\documentclass[useAMS, usenatbib, fleqn]{mn2e}

\usepackage{aas_macros}
\usepackage{url}
\usepackage{amsmath}
\usepackage{amsfonts}
\usepackage{amssymb}
\usepackage{graphicx}
\usepackage{subfig}
\usepackage{xspace}
\usepackage{float}
\usepackage{color}
\usepackage[breaklinks, colorlinks, citecolor=blue, linkcolor=black, urlcolor=black]{hyperref}
\usepackage{txfonts}

\newcommand{\ie}{{{i.e.}~}}
\newcommand{\eg}{{{e.g.}~}}
\newcommand{\equref}[1]{{\xspace}Eq.~(\ref{#1})}
\newcommand{\figref}[1]{{\xspace}Fig.~\ref{#1}}

\newcommand{\figrefbegin}[1]{{\xspace}Figure~\ref{#1}}
\newcommand{\secref}[1]{{\xspace}Sec.~\ref{#1}}

\renewcommand{\d}{{\mathrm{d}}}
\newcommand{\equ}[1]{\begin{equation}#1\end{equation}}
\newcommand{\eqn}[1]{\begin{eqnarray}#1\end{eqnarray}}

\newcommand{\negsp}[1]{\hspace*{-#1mm}}

\newcommand{\gal}{g}
\newcommand{\nobj}{{N_{\rm gal}}}
\newcommand{\zi}{j}
\newcommand{\mi}{k}
\newcommand{\ti}{i}

\newcommand{\nfreq}{{N_b}}
\newcommand{\zimin}{z_{\zi, \min}}
\newcommand{\zimax}{z_{\zi, \max}}
\newcommand{\mimin}{m_{\mi, \min}}
\newcommand{\mimax}{m_{\mi, \max}}

\pubyear{2016}
\def\LaTeX{L\kern-.36em\raise.3ex\hbox{a}\kern-.15em
    T\kern-.1667em\lower.7ex\hbox{E}\kern-.125emX}

\title[Hierarchical inference of galaxy redshift distributions]
{Hierarchical Bayesian inference of galaxy redshift distributions from photometric surveys}

\author[Leistedt, Mortlock \& Peiris]{
Boris~Leistedt,$^{1,2}$\thanks{Contact email: boris.leistedt@nyu.edu}
 Daniel~J.~Mortlock$^{3,4}$ 
and Hiranya~V.~Peiris$^{2}$\medskip\\
$^{1}$ Center for Cosmology and Particle Physics, Department of Physics, New York University, New York, NY 10003, USA \\
$^{2}$ Department of Physics \& Astronomy, University College London, Gower Street, London, WC1E 6BT, UK \\
  $^3$ Astrophysics Group, Imperial College London, Blackett Laboratory, Prince Consort Road, London, SW7 2AZ, UK\\
  $^4$ Department of Mathematics, Imperial College London, London, SW7 2AZ, UK}

\begin{document}

\maketitle 
\begin{abstract}
	Accurately characterizing the redshift distributions of galaxies is essential for analysing deep photometric surveys and testing cosmological models.
	We present a technique to simultaneously infer redshift distributions and individual redshifts from photometric galaxy catalogues.
	Our model constructs a piecewise constant representation (effectively a histogram) of the distribution of galaxy types and redshifts, the parameters of which are efficiently inferred from noisy photometric flux measurements. 
	This approach can be seen as a generalization of template-fitting photometric redshift methods and relies on a library of spectral templates to relate the photometric fluxes of individual galaxies to their redshifts. 
	We illustrate this technique on simulated galaxy survey data, and demonstrate that it delivers correct posterior distributions on the underlying type and redshift distributions, as well as on the individual types and redshifts of galaxies.
	We show that even with uninformative priors, large photometric errors and parameter degeneracies, the redshift and type distributions can be recovered robustly thanks to the hierarchical nature of the model, which is not possible with common photometric redshift estimation techniques.
	As a result, redshift uncertainties can be fully propagated in cosmological analyses for the first time, fulfilling an essential requirement for the current and future generations of surveys.
\end{abstract}

\section{Introduction}

	Testing cosmological models using the distribution of galaxies has become a routine operation thanks to large galaxy surveys such as the Sloan Digital Sky Survey (SDSS, \citealt{Gunn:2006tw}) and the Canada-France-Hawaii Telescope Lensing Survey \citep[CFHTLenS][]{Heymans:2012gg}. 
	Ongoing and upcoming imaging surveys, for example the Dark Energy Survey \citep[DES, 2012--,][]{Abbott:2005bi} and the Large Synoptic Survey Telescope \citep[LSST, 2020--,][]{Abell:2009aa} will probe even larger volumes and will allow us to constrain the large scale properties of the Universe at unprecedented accuracy. 
	This will prove essential for testing our understanding of gravity, dark matter and dark energy, as well as for searching for new physics in unexplored regimes \citep[\eg][]{Peacock:2006kj, Albrecht:2006um, Weinberg:2012es}. 
	
	The redshift distributions of galaxies are essential ingredients for exploiting photometric survey data.
	They are needed to confront measurements of the clustering and cosmic shear of galaxies with theoretical predictions.
	The accuracy of these distributions is critical since any mischaracterization can translate into significant biases in the cosmological parameters inferred from data \citep[see \eg][]{Hildebrandt:2011hb, Benjamin:2012qp, Cunha:2011qe, Huterer:2012zs, Bonnett:2015pww}.  
	Currently, all methods for estimating redshift distributions rely on the availability of redshift estimates for all galaxies in the sample of interest. 
	This is problematic because such estimates can suffer from systematic errors or increased sensitivity to prior information and training data, which significantly affects the quality of the redshift distribution estimates \citep[\eg][]{Hildebrandt:2011hb, Sanchez:2014zgq}.  
	The method proposed in this paper addresses these issues by simultaneously inferring the redshift distributions as well as individual redshift estimates.
	
	Three broad classes of methods are available to estimate redshifts from noisy photometric fluxes or magnitudes: template-fitting, machine-learning, and clustering methods. 
	Template-fitting methods  \citep[\eg][]{Benitez:1998br, Feldmann:2006wg, Brammer:2008qv} assume that each galaxy belongs to a type whose rest-frame luminosity density (simply called `spectral energy density' or `spectrum' below) is known. 
	It can be taken from a library of spectral templates constructed from data or simulations. 
	Redshift estimates are then obtained by predicting photometric fluxes as a function of redshift and comparing them with the observed fluxes. 
	The second widespread class of methods involves fitting the flux-redshift relation directly using a flexible model, via machine learning techniques such as neural networks or decision trees \citep[\eg][]{Collister:2003cz, Kind:2013eka, Sadeh:2015lsa}. 
	The third approach, sometimes referred to as `clustering redshifts', delivers redshift estimates by exploiting three-dimensional spatial information, \eg via reconstruction of the three-dimensional density field or cross-correlations of various galaxy catalogues \citep[\eg][]{Matthews:2010an, Jasche:2011qm, Choi:2015mnp}.
	The lack of training and validation data at high redshift and faint magnitudes (where most of the cosmological information is) significantly hinders these three approaches. 
	In addition, photometric calibration and distortions due to instrumental effects or observing conditions affect the quality of the measured fluxes, and hence the redshift estimates \citep[\eg][]{Leistedt:2015kka}. 
	For these reasons, photometric redshift estimation is a major challenge in the exploitation of photometric surveys. 
	Recent comparison of methods and requirements for modern surveys can be found in \eg \cite{Hildebrandt:2010hn, Cunha:2011qe, Newman:2013cac, Sanchez:2014zgq, Abate:2014cla, Schmidt:2014ela}.
	
	Recent surveys such as CFHTLenS and DES obtained redshift distributions by stacking the individual redshift posterior distributions obtained with some of the methods described above \citep{Hildebrandt:2011hb, Bonnett:2015pww}. 
	Despite its simplicity, this approach does not yield uncertainties on the redshift distribution estimates.
	Nevertheless, it is becoming increasingly important to propagate redshift uncertainties in cosmological analyses \citep[as in the recent analysis of DES Science Verification data, ][]{Abbott:2015swa}.
	In this paper, we present a technique for inferring redshift distributions through a hierarchical Bayesian model which provides full posterior distributions on the redshift distributions as well as on the individual redshift estimates.
	This approach can be seen as a generalization of template-fitting methods to the estimation of redshift distributions.
	We use the library of templates from \cite{Coleman:1980ej} and the type-redshift-magnitude likelihood function implemented in the BPZ method \citep{Benitez:1998br}.
	BPZ was one of the main codes used to produce redshift distributions estimates through stacking of the individual redshifts for CFHTLenS and DES \citep{Bonnett:2015pww, Hildebrandt:2011hb}.
	As highlighted above, the common stacking approach is highly sensitive to the priors on the redshift and type distributions, which are usually assumed to have restrictive analytical forms and calibrated on training data shallower that the full samples of interest.
	Our model relaxes these assumptions in several ways since it infers redshift distributions from the full data set under consideration using a hierarchical probabilistic model. 
	The posterior uncertainties in the individual redshift estimates are shrunk thanks to the addition of distribution information, and this shrinkage also includes the uncertainties in these distributions.
		
	We describe our model for the redshift distributions in \secref{sec:popmodel}, and describe its mapping onto photometric survey data in \secref{sec:photoinference}. 
	We extend our formalism to support arbitrary selection effects, including tomographic redshift bins, in \secref{sec:tomography}. 
	The framework is demonstrated on simulations in \secref{sec:demo}. 
	We conclude in \secref{sec:conclusion}.

\section{Population model}\label{sec:popmodel}

	Let us consider a catalogue of galaxies with intrinsic properties type $t$, redshift $z$, and apparent magnitude $m$ in some reference photometric band.
	The properties of these galaxies are assumed to be drawn from a unit-normalised distribution $p(t, z, m | \textrm{galaxy}, \textrm{survey})$, where the form of the distribution is explicitly dependent on
	the abundance of the sources as a function of redshift and intrinsic properties (as determined by \eg their luminosity function and 
	the comoving volume per unit redshift),
	and the characteristics and selection effects of the survey under consideration.
	For concision we adopt the more compact notation $p(t, z, m)$ for this distribution, but it is important to keep in mind that it depends on particularly survey-specific effects, a point which is explored further in \secref{sec:tomography}.

\begin{figure} 
\hspace*{-2mm}\includegraphics[width=8.9cm]{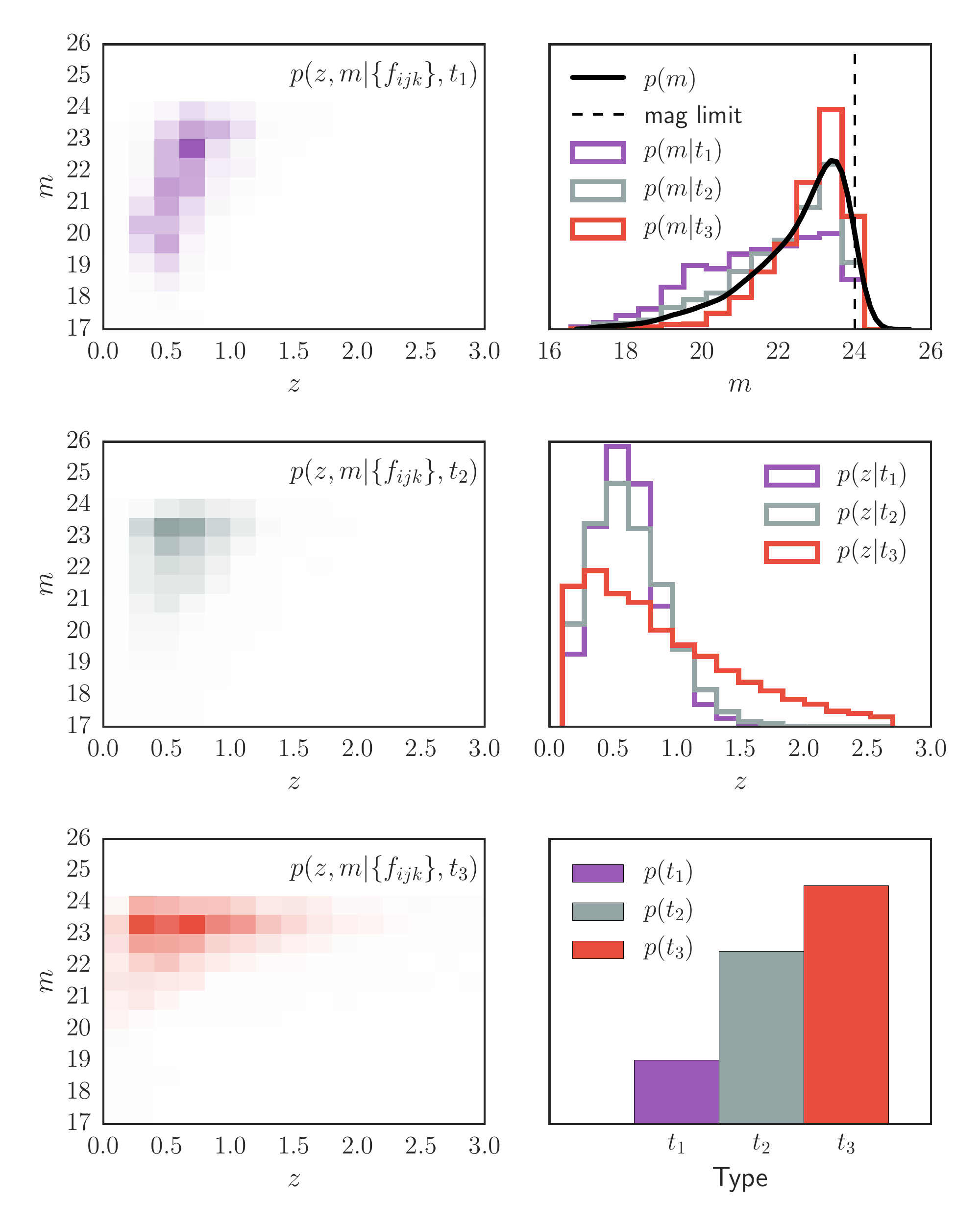}\vspace*{-3mm}
\caption{Marginal and conditional distributions describing the binned versions of the target distribution $p(t, z,m)$, parametrized by $\{f_{\ti\zi\mi}\}$. We aim to estimate these distributions using noisy photometric observations.}\label{fig:tdhistogram}
\end{figure} 

	Rather than attempting to model either the galaxy population or the observational selection effects in detail, we adopt a piece-wise constant representation of $p(t, z, m)$, parametrized by a set of coefficients $\{f_{\ti\zi\mi}\}$, such that
\[
	p(t, z,m | \{f_{\ti\zi\mi}\}) = \sum_{\ti\zi\mi} \frac{f_{\ti\zi\mi}}{(\zimax-\zimin)\,(\mimax-\mimin)} \\ 
\]
\vspace*{-3mm}
\begin{equation}
\label{eq:model_pwc}
\,\,\,
\times 
\,\,\,
\delta^\textrm{K}_{t,t_\ti} \,
\Theta(z-\zimin) \, \Theta(\zimax-z) \,
\Theta(m-\mimin) \, \Theta(\mimax-m) ,
\end{equation}
	where $\delta^\textrm{K}$ denotes Kronecker's delta and $\Theta$ the Heaviside step function. 
	This corresponds to a mathematical description of a 3-dimensional histogram, where the probability of finding an object in the bin labelled $\ti\zi\mi$ is 
$f_{\ti\zi\mi}$. 
	The type-, redshift- and magnitude-bins are indexed with $\ti = 1, \dots, N_t$, $\zi = 1, \dots, N_z$, and $\mi = 1, \dots, N_m$, respectively. 
	The redshift bins have bounds $( \zimin, \zimax )$ and the magnitude bins $( \mimin, \mimax )$. 
	This model does not formally require the bins to be contiguous or of equal size, but they should not overlap. 
	In practice they should simply tile the redshift and magnitude ranges of interest and we will adopt contiguous, equal-size, bins so that the histogram can be interpreted as a piecewise constant approximation of $p(t, z, m)$, such that
\eqn{
	f_{\ti\zi\mi} \equiv \int_{\zimin}^{\zimax}  \int_{\mimin}^{\mimax} \ p(t_\ti, z, m ) \ \d z \ \d m .
}
	
\begin{figure} 
\includegraphics[width=8.5cm]{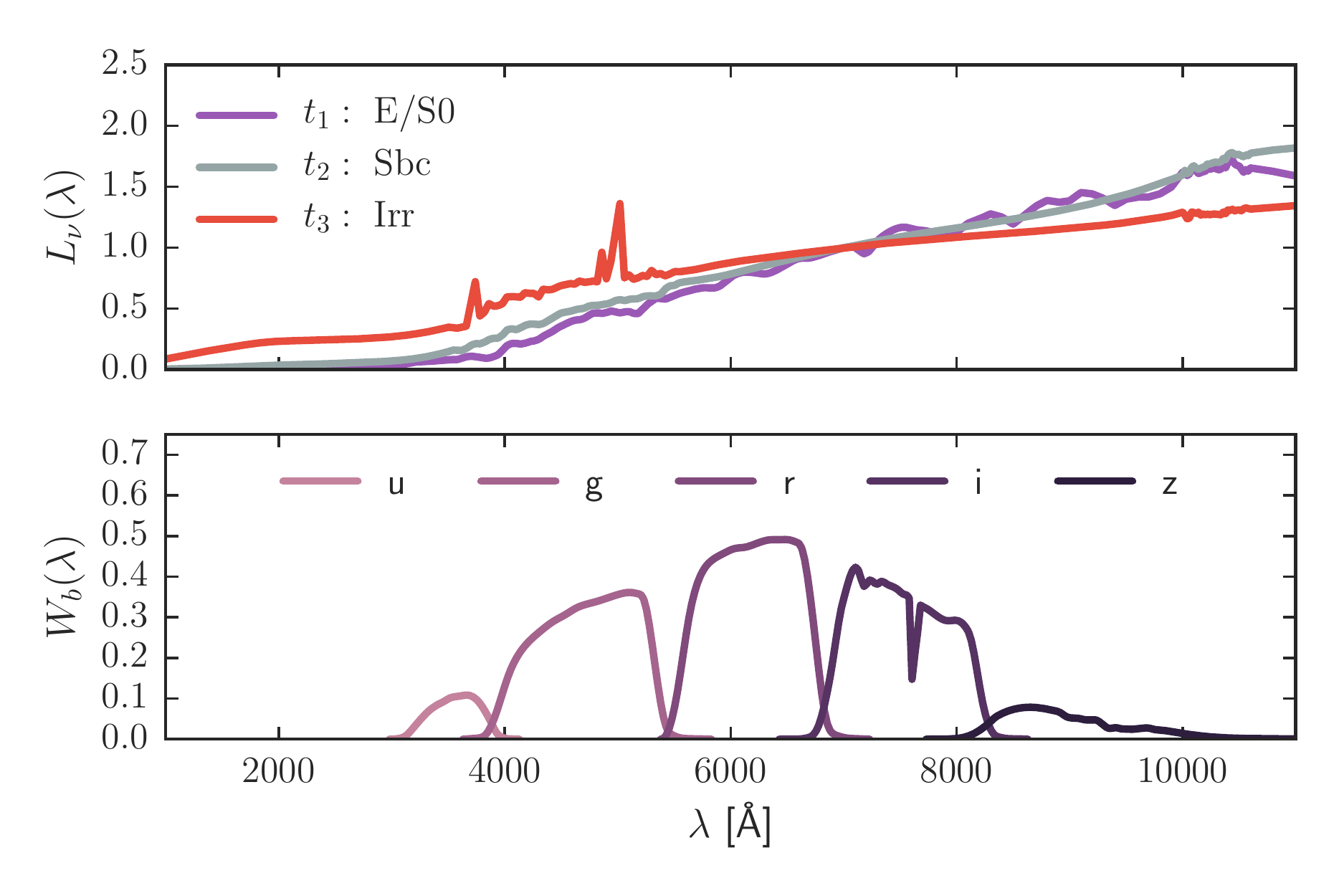}\vspace*{-3mm}
\caption{Galaxy spectral energy distributions and filter response curves used in this work. The spectral templates are taken from \citet{Coleman:1980ej} and are  normalised to $1$ at $\lambda=7500$ \AA. The filters are the SDSS $ugriz$ photometric bands \citep{Fukugita:1996qt}. 
}\label{fig:bandsandtemplates}
\end{figure}

	An example of this 3D histogram representation with $N_t=3$ and $N_z=N_m=15$ is shown in \figref{fig:tdhistogram}. 
	It is constructed from the priors $p(t, z, m)$ implemented in the BPZ code, which are parametrized with $p(t|m) = f_t \exp[-k_t (m-20)]$ and $p(z|t,m) \propto z^{\alpha_t} \exp\{-[z/z_{mt}(m)]^{\alpha_t}\}$. 
	The parameters $f_t, k_t, z_{mt}$ and $\alpha_t$ are taken from Table 1 of \cite{Benitez:1998br} and correspond to the E/S0, Sbc, and Irr templates from \cite{Coleman:1980ej}, shown in \figref{fig:bandsandtemplates}.
	In what follows we will sometimes refer to the type-redshift and redshift distributions (with other parameters marginalized over); these can be written as $p(t, z| \{\sum_\mi f_{\ti\zi\mi}\})$ and $p(z | \{\sum_{\ti\mi} f_{\ti\zi\mi} \})$, respectively.

\begin{figure} 
\hspace*{-3mm}\includegraphics[width=9.cm]{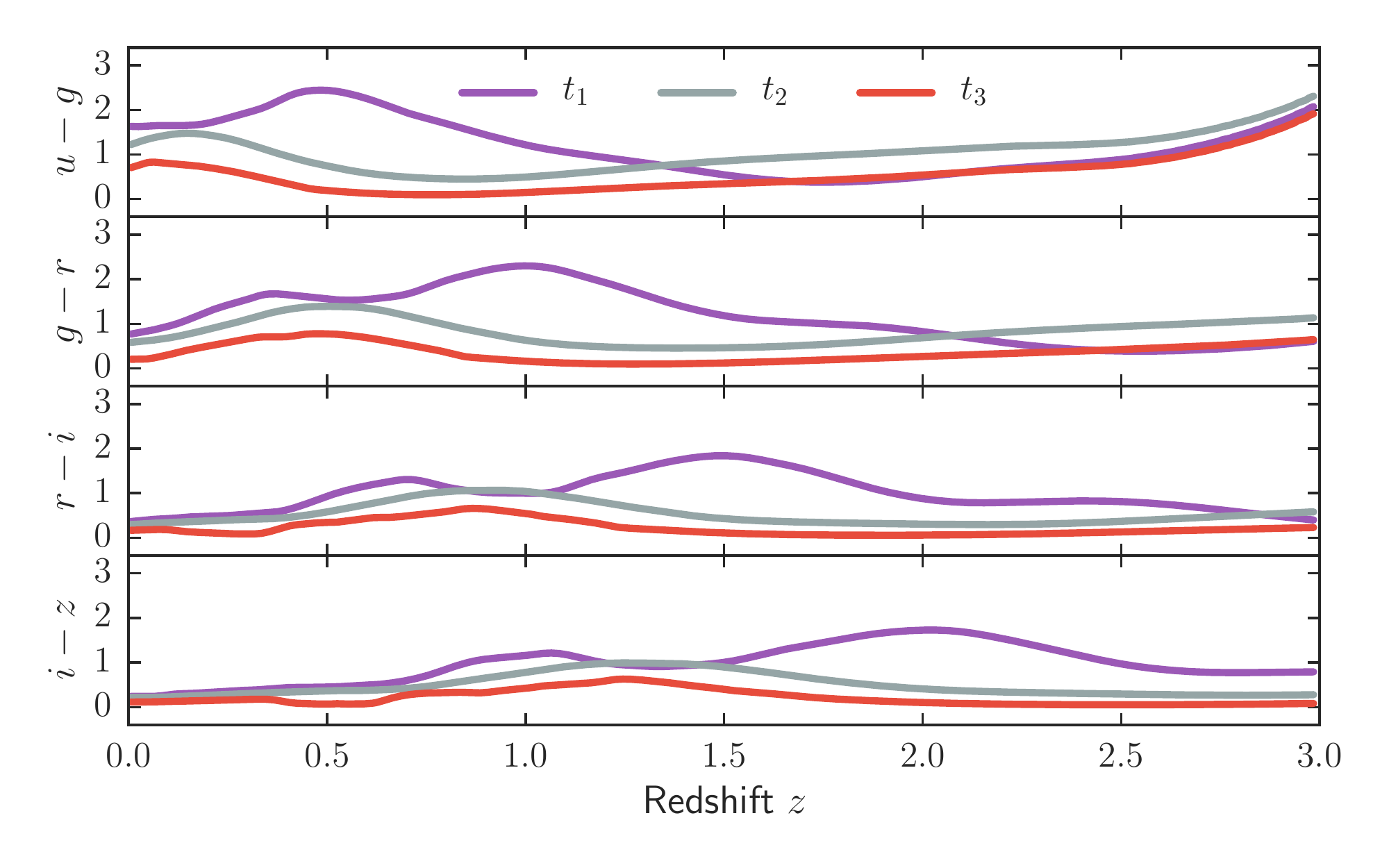}\vspace*{-3mm}
\caption{Tracks of the three templates shown in \figref{fig:bandsandtemplates} in the space of SDSS colours as a function of redshift.}\label{fig:colourtracks}
\end{figure} 

\begin{figure} 
\centering
\includegraphics[trim={0 0.7cm 0 0.2cm}, clip, width=6cm]{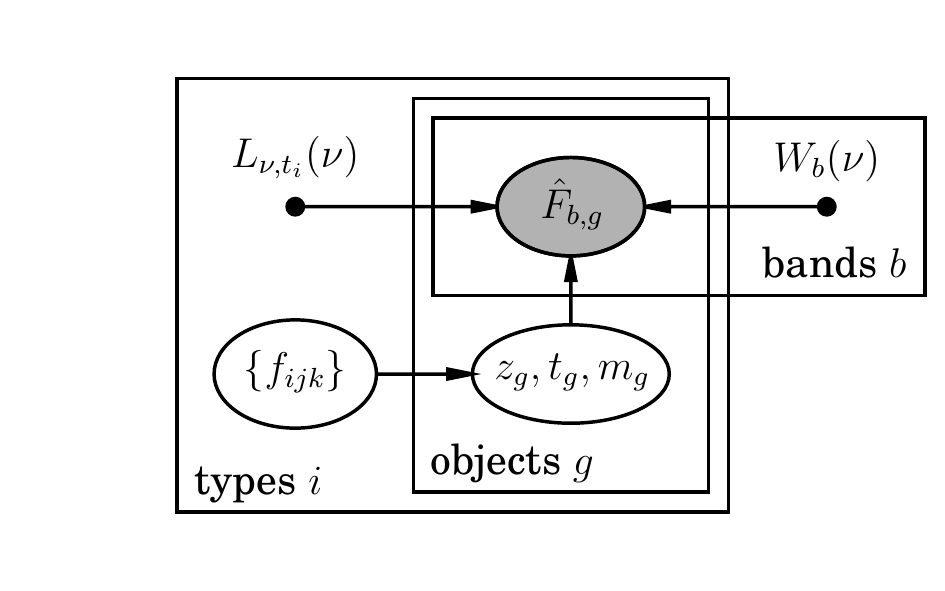}
\caption{Graphical representation of the hierarchical model, following the notation summarised in Table~\ref{tab:notation}. Dots, circles and shaded circles indicate fixed quantities, parameters to be inferred, and observed quantities, respectively. Arrows express dependency and boxes replication. $\nobj$ galaxies with fluxes measured in $\nfreq$ photometric bands are modelled using a set of $N_t$ spectral templates $L_{\nu, t_i}(\nu)$. The distribution of the intrinsic galaxy properties (type, redshift, and magnitude in the reference band) is parametrized with $\{f_{\ti\zi\mi}\}$.}\label{fig:model}
\end{figure} 

\begin{table} 
\centering
\begin{tabular}{cl}
\hline
$z$	&	redshift \\
$m$	&	apparent magnitude in the reference band \\
$\ti$	&	index of galaxy type $t_i$ \\
$\zi$	&	index of redshift bin with bounds $( \zimin, \zimax )$   \\
$\mi$	&	index of magnitude bin with bounds $( \mimin, \mimax )$ \\
$N_t$	&	number of types 	\\
$N_z$	&	number of redshift bins 	\\
$N_m$	&	number of magnitude bins 	\\
$\gal$	&	index of galaxy \\
$\nobj$	&	total number of galaxies in the sample	\\
$n_{\ti\zi\mi}$	& 	galaxy count in the $\ti\zi\mi$-th type-redshift-magnitude bin  \\
$\{n_{\ti\zi\mi}\}$	&	set of all galaxy counts $n_{\ti\zi\mi}$, summing to $\nobj$\\
$f_{\ti\zi\mi}$	&	fractional galaxy count in the $\ti\zi\mi$-th bin  \\
$\{f_{\ti\zi\mi}\}$	&	set of all fractional bin counts $f_{\ti\zi\mi}$, summing to $1$\\
$z_\gal, t_\gal, m_\gal$	&	properties of the $\gal$-th galaxy	\\
$\{z_\gal, t_\gal, m_\gal\}$	&	set of properties of all galaxies in the sample	\\
$L_{\nu, t}$	&	spectral template of the type $t$ 	\\
$b$ &	index of photometric band	\\
$\nfreq$	&	number of photometric bands	\\
$W_b$	&	$b$-th photometric filter 	\\
$F_b(t, z)$	& noiseless photometric flux (type $t$, redshift $z$, band $b$) \\
$\{\hat{F}_b\}_\gal$	& $\nfreq$ observed photometric fluxes of the $\gal$-th galaxy \\
$\{\hat{F}_{b,\gal}\}$	& observed photometric fluxes for all $\nobj$ galaxies \\
\hline
\end{tabular}
\caption{Summary of our notation. }
\label{tab:notation}
\end{table} 

\section{Inference methodology}\label{sec:photoinference}

	We now turn to the problem of inferring the parameters $\{f_{\ti\zi\mi}\}$ from a set of $\nobj$ galaxies, the properties of which are denoted by $\{ t_\gal, z_\gal, m_\gal\}$ with $\gal \in \{1, \ldots, \nobj\}$. 
	We first consider the `noiseless' case where these properties are assumed to be known for all objects (\secref{sec:noiseless}) before treating the more general case where the individual galaxies' properties  must be inferred simultaneously with the distributions of interest from noisy photometric data (\secref{sec:noisyinference}).
	In both cases we use a uniform (maximally uninformative) prior on the coefficients $\{f_{\ti\zi\mi}\}$, \ie subject to the constraints  $ 0\leq f_{\ti\zi\mi} \leq1$ for all $\ti\zi\mi$ and $\sum_{\ti\zi\mi} f_{\ti\zi\mi}=1$, which is a Dirichlet distribution, given by
\equ{
\label{eq:prior}
p(\{f_{ijk}\} ) =
  (N_t\, N_z\, N_m -  1 )! \,
  \delta_\textrm{D}\left(1-\sum_{\ti\zi\mi} f_{\ti\zi\mi}\right)
  \, \prod_{\ti=1}^{N_t}\prod_{\zi=1}^{N_z}\prod_{\mi=1}^{N_m}
  \Theta(f_{ijk}),
}
where $\delta_\textrm{D}(x)$ is the Dirac delta function.
	A summary of our notation is provided in Table~\ref{tab:notation}.

\subsection{Noiseless case}\label{sec:noiseless}

	In the noiseless case a set of sufficient statistics for the full list of galaxy properties, $ \{ t_\gal, z_\gal, m_\gal\}$, is the numbers in each of the type, redshift and magnitude bins, which is given by
\equ{\begin{split}\label{eq:nbcounts}
        n_{\ti\zi\mi} = \sum_\gal \  \delta^\textrm{K}_{t_\gal,t_\ti} &\, \Theta(z_\gal-\zimin)  \, \Theta(\zimax-z_\gal)  \\
        &\times \ \Theta(m_\gal-\mimin) \, \Theta(\mimax-m_\gal).
\end{split}}
The likelihood of the binned data is given by a multinomial distribution,
\equ{
p(\{n_{\ti\zi\mi}\} | \{f_{\ti\zi\mi}\})
  = \nobj! \, 
  \prod_{\ti=1}^{N_t}\prod_{\zi=1}^{N_z}\prod_{\mi=1}^{N_m} 
  \frac{ f_{\ti\zi\mi}^{n_{\ti\zi\mi}} }{ n_{\ti\zi\mi}! } .
}
Combining the above prior and likelihood leads to 
a posterior that, like the prior, is a Dirichlet distribution,
which reads
\equ{\label{eq:conditional1}
p(\{f_{\ti\zi\mi}\} | \{ n_{\ti\zi\mi} \})  
}
\[
 = 
( \nobj + N_t \, N_z \, N_m -  1 )! \,
\delta_\textrm{D}
\left(1-\sum_{\ti\zi\mi} f_{\ti\zi\mi}\right)  \,
     \prod_{\ti=1}^{N_t}\prod_{\zi=1}^{N_z}\prod_{\mi=1}^{N_m} \frac{ \Theta(f_{\ti\zi\mi})\, f_{\ti\zi\mi}^{n_{\ti\zi\mi}} }{ n_{\ti\zi\mi}! } .
\]
	Thus, inferring the parameters $\{f_{\ti\zi\mi}\}$ when the types, redshifts and magnitudes of the sources are known is possible, thanks to the analytic posterior distribution that only requires the bin counts $\{ n_{\ti\zi\mi} \}$. 
	Before we map this formalism to photometric galaxy survey observations, we make three important remarks.
	
	First, in the limit of uncorrelated bins and large bin counts $\{n_{\ti\zi\mi}\}$, the marginalized mean and variance on $f_{\ti\zi\mi}$ both reduce to $n_{\ti\zi\mi}/\nobj$, which is the classical approximate histogram estimator.  
	However, this estimator fails in the regime of strong inter-bin correlations or quasi-empty bins. This is illustrated in Appendix~\ref{app:gaussiancomp}.
	
	Second, the equations of this section imply an important property of the Dirichlet model: it is agnostic to the nature of the bins, \ie to their size or physical interpretation. 
	These are only needed for the calculation of the number counts $n_{\ti\zi\mi}$. 
	This is a manifestation of the categorical nature of the model: objects drawn from $N$ categories follow a multinomial distribution, and the posterior distribution on its parameters is a Dirichlet distribution. 
	In practice, the only consequence of this property is the insensitivity of the model to the nature of the binning or the order of the labelling $ijk$, as all the previous equations could have been casted with a unique categorical label $\ell\equiv ijk$. The Dirichlet posterior would then read $p(\{f_{\ell}\} | \{ n_{\ell} \})\propto \prod_{\ell} { f_{\ell}^{n_{\ell}} }/{ n_{\ell}! }$. Yet, the explicit 3-dimensional formalism above will prove essential for the clarity of \secref{sec:noisyinference}, where types, redshifts and magnitudes are distinct physical quantities in the hierarchical model, and are marginalized over separately.
	
	Third, prior knowledge about the correlations between bins can be incorporated by considering a prior $p(\{f_{\ti\zi\mi}\})$ following a generalised Dirichlet distribution. 
	In this case, the posterior distribution on $\{f_{\ti\zi\mi}\}$ is also a generalised Dirichlet distribution, and the model is no longer agnostic to the nature and order of the bins. 
	We do not consider this case in this paper but we highlight that this extension does not affect the methods presented below.
	In particular, in the limit of a large number of galaxies or weak priors on the inter-bin correlations the generalised Dirichlet distribution reduces to a standard Dirichlet case.

\subsection{Inference using photometric data}\label{sec:noisyinference}

	In this section we no longer assume that the types, redshifts and magnitudes of the objects are known. These must now be inferred simultaneously with the underlying distributions, using only noisy photometric observations.

\subsubsection{Likelihood}

	In imaging surveys, the main observable
	\footnote{Morphological information about sources is also potentially important: it enables numerous cosmological studies and improves the quality of star-galaxy separation and redshift estimation. 
	However, we focus on magnitude information here in order to exploit widely-used libraries of spectral templates, which typically do not use morphological information.} 
	is the photon flux measured in $\nfreq$ bands indexed by $b=1, \dots, \nfreq$, 
\equ{
	F_b(t, z) =  \frac{1+z}{4\pi D_L^2(z)} \int_{0}^{\infty} \frac{{\rm d} \nu}{\nu}\ 
{L_{\nu, t}\left[\nu({1+z})\right]}\ W_b(\nu),
}
 	 where $L_{\nu, t}(\nu)$ is the rest-frame luminosity density of the extragalactic source of type $t$ as a function of frequency $\nu$ \citep[see \eg][]{Hogg:2002yh}. 
	$W_b(\nu)$ are functions characterizing the response of the photometric filters. 
	\figrefbegin{fig:bandsandtemplates} shows the three spectral templates as well as the photometric filters used in this paper. 
	The templates are the E/S0, Sbc, and Irr templates from \cite{Coleman:1980ej} as packaged by BPZ \citep{Benitez:1998br}, while the filters are the SDSS $ugriz$ photometric bands \citep{Fukugita:1996qt} including the effects of extinction and airmass\footnote{\url{http://classic.sdss.org/dr7/instruments/imager}}.  
	Magnitudes in the AB system are related to flux densities via
	\equ{
	m_b(t,z)=-2.5 \log \left[ \frac{F_b(t,z)}{\int  g_\nu(\nu) \, W_b(\nu) \, {{\rm d} \nu}/{\nu}}\right],
}
	with $g_\nu(\nu) = 3631$ Jy.
	Finally, we take the reference magnitude $m$ to be the $i$ band magnitude. 
	
	For a given galaxy $g$, noisy measurements of its photometric fluxes are available,
\eqn{
	\{\hat{F}_{b}\}_\gal=(\hat{F}_{1,\gal}, \dots, \hat{F}_{\nfreq,\gal}),
}
	with errors $\sigma_{\hat{F}_{1,\gal}}, \dots, \sigma_{\hat{F}_{\nfreq,\gal}}$. 
	We define a multidimensional Gaussian likelihood function in flux space,
\eqn{
	p( \{\hat{F}_{b}\}_\gal| t, z, m ) = \prod_{b=1}^{\nfreq} \mathcal{N}\left[F_b(t,z) ; a\hat{F}_{b,\gal}, \sigma^2_{\hat{F}_{b,\gal}}\right],
}
	where $\mathcal{N}(x; \mu, \sigma^2)$ denotes the standard normal distribution of mean $\mu$ and variance $\sigma^2$, evaluated at $x$.
	Note that this likelihood does not depend on $m$, and that $a$ is the (arbitrary) template normalization, which we marginalize over analytically following \cite{Benitez:1998br}.
	A different form could be adopted (\eg in colour space, provided the correlations between colours sharing a common band was incorporated), with no impact on the methods and conclusions of this paper. 

	As each spectral template $L_{\nu,t}$ is redshifted and integrated in the photometric bands, it forms a set of tracks in flux-redshift or colour-redshift space, shown in \figref{fig:colourtracks} for the templates and filters considered here.

\begin{figure*} 
\hspace*{-2mm}\includegraphics[width=18cm]{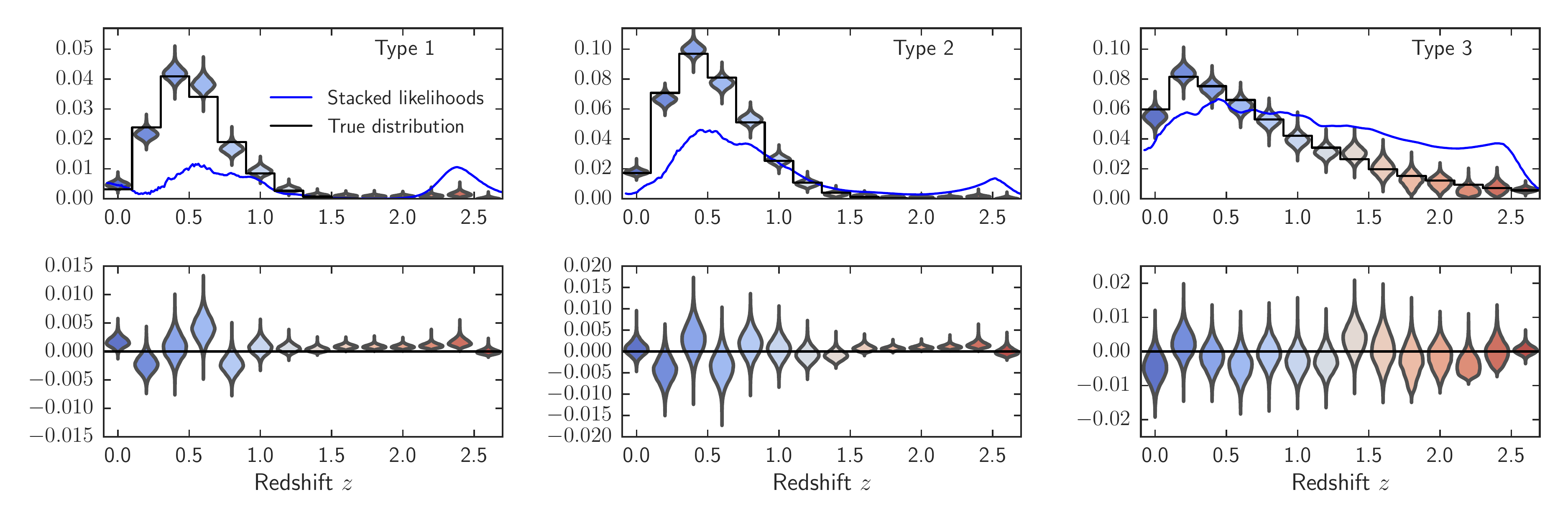}\vspace*{-3mm}
\caption{Type-redshift distributions with marginalization over the reference magnitude. The solid lines show the input true distributions while the distributions obtained with the inferred parameters $f_{\ti\zi\mi}$ are shown as violin plots, \ie box plots whose profiles represent the distribution of parameter values. The bottom panels show the difference and demonstrate that the model not only recovers the input distributions but also provides meaningful error bars. This is illustrated in further detail in \figref{fig:dirichletparamsposterior}.}
\label{fig:recovereddistributions}
\end{figure*} 

\begin{figure*} 
\centering
\includegraphics[trim={5.5cm 2.4cm 5.5cm 2.2cm}, clip, width=16cm]{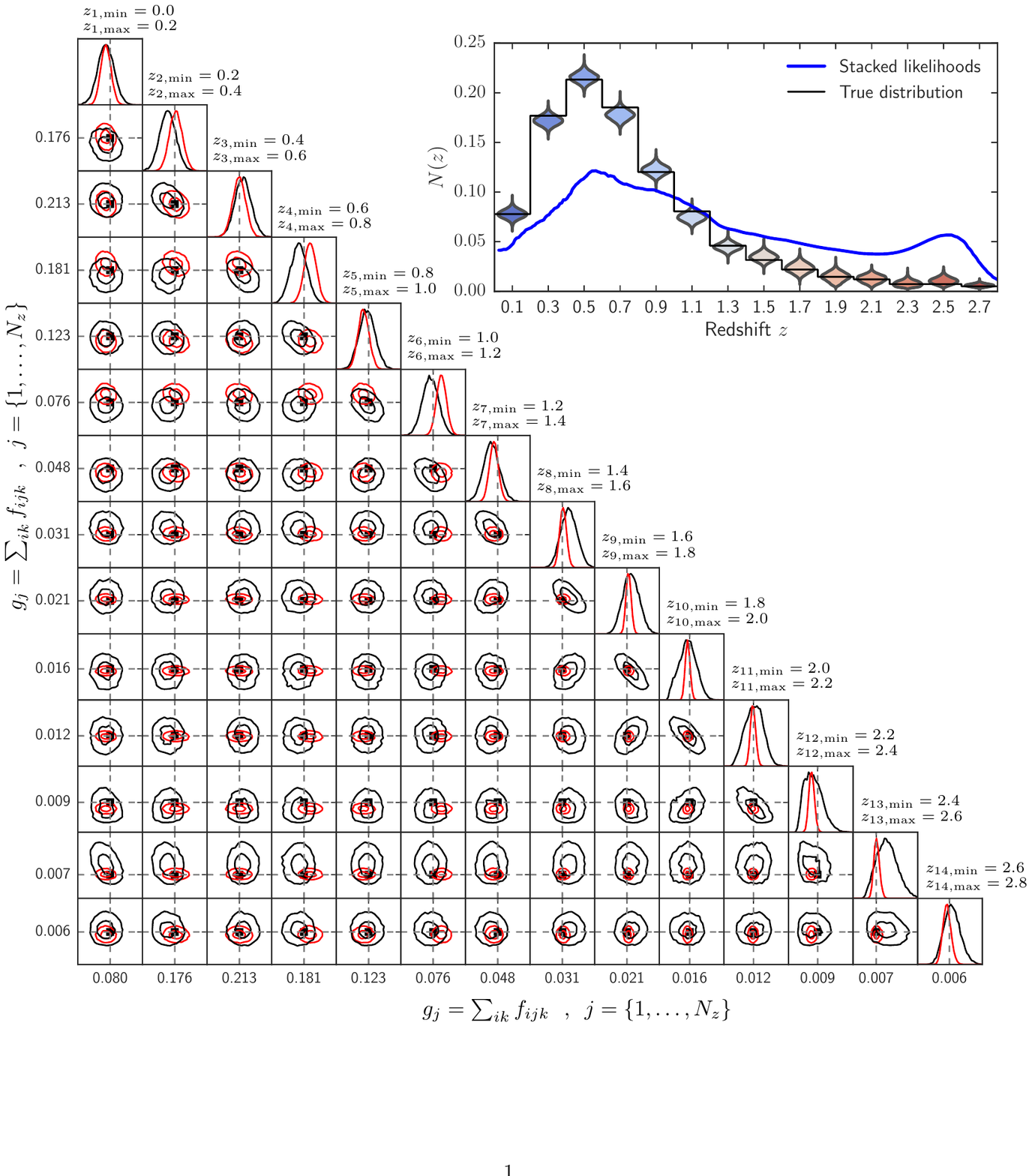}
\caption{Posterior distributions on the fractional count parameters $g_\zi = \sum_{\ti\mi} f_{\ti\zi\mi}$, \ie on the type-redshift-magnitude parameters $f_{\ti\zi\mi}$ with marginalization over type and magnitude. The top right panel shows the resulting redshift distributions, with the violin plots illustrating how parameter uncertainties propagate into the redshift distribution estimates.  Adjacent bins are significantly anticorrelated, as expected from a Dirichlet model properly harnessing the individual likelihoods. The red contours show the posterior distributions on the same parameters from a noise-free case, \ie a standard Dirichlet model where the types, redshifts and magnitudes of all objects are known.}\label{fig:dirichletparamsposterior}
\end{figure*} 

\subsubsection{Efficient parameter inference}

	The population model and the flux-redshift likelihood can be combined into the hierarchical model shown in \figref{fig:model}. 
	The observed quantities of this model are the set of $\nfreq \times \nobj$ fluxes, denoted by $\{ \hat{F}_{b,\gal}\}$. The parameters of interest are the $\nobj$ triplets of intrinsic parameters $\{ t_\gal, z_\gal, m_\gal\}$ as well as the population parameters $ \{f_{\ti\zi\mi}\}$. 
	The full, joint posterior distribution on these parameters reads
\eqn{
	&& \negsp{6} p(\{ t_\gal, z_\gal, m_\gal\}, \{f_{\ti\zi\mi}\} \ | \ \{ \hat{F}_{b,\gal}\} ) \   \nonumber  \\
	&& \propto \negsp{2}\quad p( \{f_{\ti\zi\mi}\} )  \prod_{g=1}^{\nobj}  p(\{\hat{F}_{b}\}_\gal| z_\gal, t_\gal, m_\gal) \ p(z_\gal, t_\gal, m_\gal | \{f_{\ti\zi\mi}\} ) .
}
	Unlike the noise-free case of the previous section, the implied normalised posterior does not have an analytic form. 
	It is possible to directly draw samples from this posterior using a two-step Gibbs sampler because the conditional posterior distributions can be easily sampled. 
	At a given Gibbs iteration, a sample of $\{f_{\ti\zi\mi}\}$ is drawn from $p(\{f_{\ti\zi\mi}\} | \{ t_\gal, z_\gal, m_\gal\}, \{ \hat{F}_{b,\gal}\} )$. 
	The latter follows the Dirichlet model of \equref{eq:conditional1}, where the number counts $\{n_{\ti\zi\mi}\}$ are calculated from the triplets $\{ t_\gal, z_\gal, m_\gal\}$ of the previous Gibbs iteration. 
	Then, $\{ t_\gal, z_\gal, m_\gal\}$ are updated using the newly drawn $\{f_{\ti\zi\mi}\}$ by looping over galaxies and updating each triplet $(t_\gal, z_\gal, m_\gal)$ using
\equ{\begin{split}
	p( t,  z, m | &\{f_{\ti\zi\mi}\},\{\hat{F}_{b}\}_\gal )   \\
	= \ \sum_{\ti\zi\mi} &\quad \frac{f_{\ti\zi\mi}  \times p( \{\hat{F}_{b}\}_\gal| t, z, m )  }{(\zimax-\zimin)(\mimax-\mimin)} \\
	& \quad \times \delta^\textrm{K}_{t,t_\ti} \\ 
	& \quad \times \Theta(z-\zimin)  \Theta(\zimax-z) \\
	& \quad \times  \Theta(m-\mimin)  \Theta(\mimax-m) .
	\end{split}
}
	As for a classical Gibbs sampler, alternately drawing $\{f_{\ti\zi\mi}\}$ and $\{ t_\gal, z_\gal, m_\gal\}$ from the previous conditional distributions allows one to explore the full joint posterior distribution of interest.
 
	If one is not interested in the properties of individual
galaxies $\{ t_\gal, z_\gal, m_\gal\}$, but only in their distributions described by $\{f_{\ti\zi\mi}\}$, then a significant speed up can be achieved by binning $p( \{\hat{F}_{b}\}_\gal| t, z, m )$ in the second step of the Gibbs sampler, \ie by using a binned likelihood
\eqn{
	p( \{\hat{F}_{b}\}_\gal| \ti\zi\mi ) =  \int_{\zimin}^{\zimax}  \int_{\mimin}^{\mimax}  \ p( \{\hat{F}_{b}\}_\gal| t_\ti, z, m ) \   \d m \ \d z .
}
	This is because the Dirichlet model only requires the bin counts $\{n_{\ti\zi\mi}\}$, and those can be updated directly by drawing bin locations $\ti\zi\mi$ from the binned likelihood $p( \{\hat{F}_{b}\}_\gal| \ti\zi\mi )$, bypassing the parameters $\{ t_\gal, z_\gal, m_\gal\}$.

\section{Selection effects, redshift tomography}\label{sec:tomography}

\subsection{Background} 

		One interesting extension of the formalism presented above is the inclusion of selection effects. 
	An example common to all photometric surveys is the existence of a detection probability characterizing  how objects have been added to the sample (including the completeness of the survey) for which the redshift distribution must be estimated. 
	In the example presented in \secref{sec:demo}, we will include this effect via a magnitude limit in the reference band (magnitude limits are not applied to the other bands).
	This is directly captured by our formalism because the reference magnitude itself is a dimension of the distribution $p(t,z,m)$ and thus parametrized by $\{f_{\ti\zi\mi}\}$. 
	However, any selection effects not directly involving type, redshift or reference magnitude will not be captured in the previous formalism. 
	
	A concrete example is the introduction of a second magnitude limit in another band, which would occur if galaxies had to be simultaneously detected in two bands in order to be retained in the sample. 
	This selection effect would naturally get imprinted in the redshift distributions in a non-trivial way (correlated with type, redshift and reference magnitude). 
	Another common example is the splitting of a main galaxy catalogue into redshift bins, which is of interest to study the redshift evolution of galaxy clustering or cosmic shear observables \citep[\eg][]{Benjamin:2012qp, Leistedt:2015kka, Crocce:2015xpb, Bonnett:2015pww}. 
	In this section we show how to deal with such survey-related selection effects.

\subsection{General framework}

	In our hierarchical model, an extra parameter per object must be introduced to support selection effects in general. 
	We denote this parameter by $s$, and without loss of generality assume that it depends on the observed fluxes $\{\hat{F}_b\}$. 
	Our distributions and observations are now conditioned on $s$, and the full posterior distribution is
\eqn{
	&& \negsp{6} p(\{ t_\gal, z_\gal, m_\gal\}, \{f_{\ti\zi\mi}\} \ | \ \{ \hat{F}_{b,\gal}\}, \{s_\gal\} ) \    \\
	&& \propto \negsp{2}\quad  \prod_{g=1}^{\nobj}  p(\{\hat{F}_{b}\}_\gal | z_\gal, t_\gal, m_\gal, s_\gal) \  p(z_\gal, t_\gal, m_\gal | \{f_{\ti\zi\mi}\}, s_\gal ) \ p( \{f_{\ti\zi\mi}\} | s_\gal ) \nonumber .
}
	The motivation for adopting this form is the following: $s$ indicates whether or not an object is included in the sample of interest. 
	All objects in the sample have the same value of $s$. 
	Therefore, we are truly interested in inferring the parameters $\{f_{\ti\zi\mi}\} $ describing $p(t,z,m|\{f_{\ti\zi\mi}\} , s)$, which is modelled as a piecewise function as above. 
	In other words, the formalism of the previous sections can be used to estimate $\{f_{\ti\zi\mi}\}$ with no other changes than using the modified likelihood function $p(\{\hat{F}_{b}\} | z, t, m, s)$. 
	The latter can be expressed in terms of our original likelihood via
\equ{
	p( \{\hat{F}_b\} | s, t, z, m ) =  \frac{  p( \{\hat{F}_b\} | t, z, m ) \ p(s | \{\hat{F}_b\}) }{ p( s |  t, z, m ) },
}
	where we have omitted the $g$ subscripts for conciseness. 
	The first term of the numerator is the original flux-redshift likelihood, while the second term describes the selection effect under consideration. 
	The denominator can be expressed as a marginalization over possible fluxes of these two terms,
\equ{
	p( s |  t, z, m ) =  \int   p(s | \{\hat{F}_b\})\ p( \{\hat{F}_b\} | t, z, m )\ {\rm d} \hat{F}_1 \ldots {\rm d}\hat{F}_\nfreq.
}

	Carrying out the inference with $p( \{\hat{F}_b\} | s, t, z, m )$ instead of $p( \{\hat{F}_b\} | t, z, m )$ will produce posterior distributions on the parameters $\{f_{\ti\zi\mi}\}$ that describe $p(t,z,m|\{f_{\ti\zi\mi}\} , s)$, the distribution of interest with the selection effect included.

\subsection{Illustration}

	In order to understand the effect on our likelihood and gain intuition we now consider a simplified example. 
	Let us assume that instead of observing fluxes $\{\hat{F}_b\}$, we were directly observing a noisy redshift estimate $\hat{z}$ with error function $p( \hat{z} | z ) = \mathcal{N}(z; \hat{z}, \sigma^2_{\hat{z}})$. 

	Further, our selection effect is to retain objects for which $\hat{z}$ is in some range $[z_{\rm min}, z_{\rm max}]$. 
	In other words we have the top hat function
\equ{
	p(s | \hat{z}) = \Theta(\hat{z}-z_{\rm min}) \, \Theta(z_{\rm max}-\hat{z}) .
}
	Using the previous expressions, our likelihood function conditioned on $s$ reads
\equ{
	p( \hat{z} | s, z ) =  \frac{ p(s | \hat{z}) \, p( \hat{z} | z ) }{ p( s |  z ) },
}
with
\equ{
	p( s |  z ) =  \int   p(s | \hat{z})\, p( \hat{z} | z ) \, {\rm d} \hat{z}.
}
	Importantly, because $ \mathcal{N}(z; \hat{z}, \sigma^2_{\hat{z}}) = \mathcal{N}(z - \hat{z}; 0, \sigma^2_{\hat{z}})$, we can also write $p( s |  z )$ as a simple convolution of the selection effect with our error function.
	In our case this is a top hat smoothed with a Gaussian error function.

	Returning to working with fluxes $\{\hat{F}_b\}$, since our likelihood is a multidimensional Gaussian the distribution $p( s |  t, z, m )$ can also be seen as a multidimensional convolution. 
	The main difference is that $p(s | \{\hat{F}_b\})$ may take a more complicated form than a top hat. 
	In particular, the case of redshift tomography can be implemented by selecting galaxies whose maximum likelihood redshift estimate, denoted by $z_{\rm ML}(\{\hat{F}_b\})$, is in a  range $[z_{\rm min}, z_{\rm max}]$. 
	In this case we have
\equ{
	p(s | \{\hat{F}_b\}) = \Theta[z_{\rm ML}(\{\hat{F}_b\})-z_{\rm min}] \ \Theta[z_{\rm max}-z_{\rm ML}(\{\hat{F}_b\})].
}
	As a result, $p( s |  t, z, m )$ is a multidimensional convolution of the Gaussian likelihood function with a complicated domain defined by $p(s | \{\hat{F}_b\})$.
	 However, depending on the form of the selection effect and the likelihood function, it might also be possible to draw from $p( \{\hat{F}_b\} | s, t, z, m )$ directly.
	 Both $p( s | t, z, m )$ and $p( \{\hat{F}_b\} | s, t, z, m )$ can be precomputed for each object prior to inferring the distribution parameters $\{f_{\ti\zi\mi}\}$.

\begin{figure} 
\includegraphics[width=8.5cm]{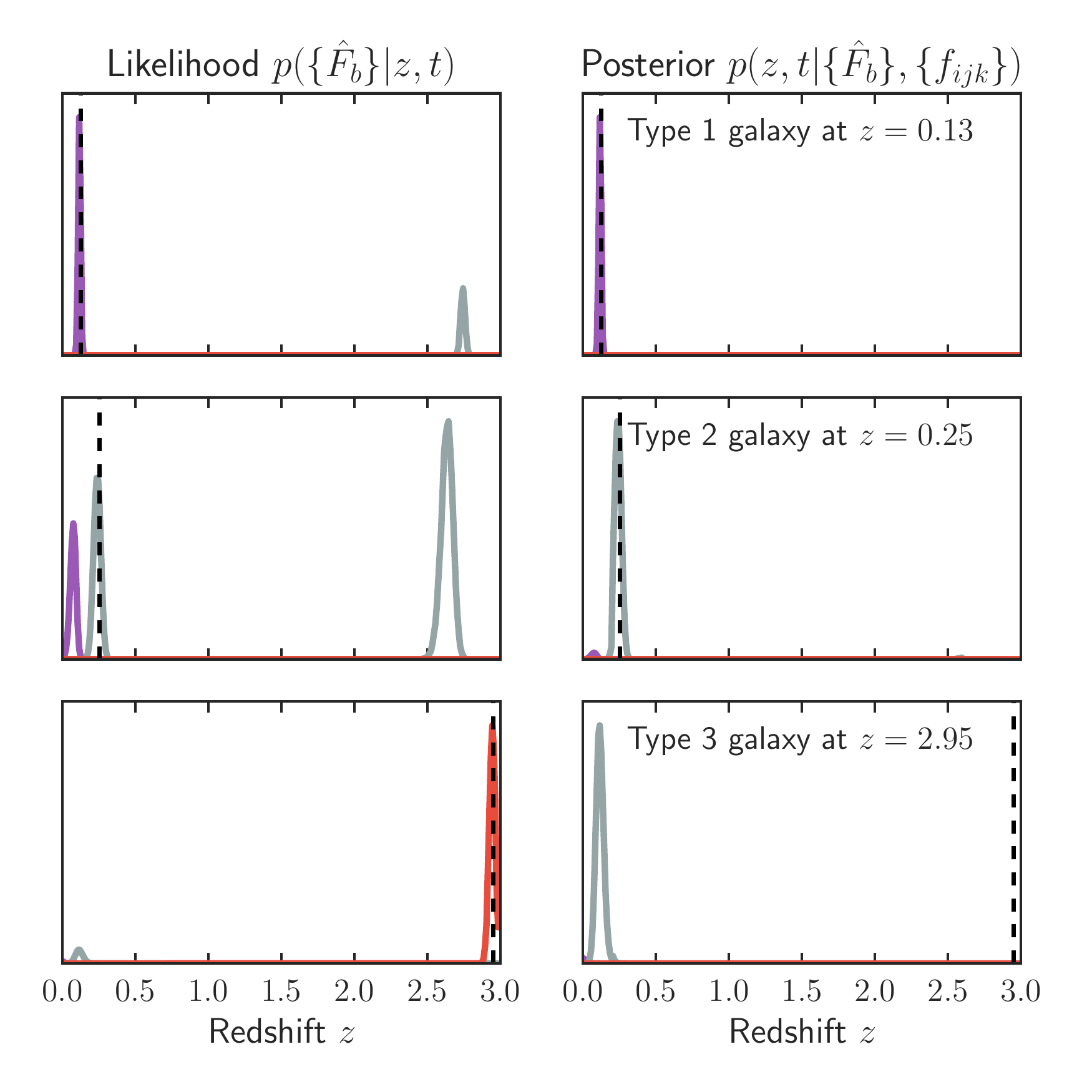}\vspace*{-3mm}
\caption{Likelihood functions and posterior distributions for three objects from our simulation -- one of each type. The dashed vertical lines show the true redshift. Inferring the individual types and redshifts as well as their underlying population removes some of the strong degeneracies and secondary peaks in the flux-redshift likelihood function.}\label{fig:objlikeandposts}
\end{figure} 

\section{Demonstration on simulations}\label{sec:demo}

	We now present an illustration of our methodology on simulations of photometric data. 
	Because they do not affect the features of our method, we adopt simple galaxy and noise distributions and do not include selection effects other than a single magnitude limit in the reference band.
	
	We start by drawing $\nobj=10^4$ $i$-band magnitudes from a realistic magnitude distribution $p(m)$ with a 5-$\sigma$ magnitude limit of $24$, which is shown in the top right panel of \figref{fig:tdhistogram}. 
	This is achieved by drawing a large number of objects according to $p(m) \propto \exp(m)$, then adding Gaussian noise and only retaining the first $\nobj$ objects detected at more than five sigma significance. 
	In this work we use the noise law of \cite{rykoff:2015ran} for all bands. The latter reads 
	\equ{
		\sigma(m) = \frac{2.5}{\ln10}\sqrt{{\left(1+\frac{F_{\rm noise}}{F(m)}\right)}\frac{1}{F(m)t_{\rm eff}}},
	} 
	with $F_{\rm noise} = t_{\rm eff}F^2(m_{\rm lim})/25 - F(m_{\rm lim})$ and $t_{\rm eff} = \exp(a+b(m_{\rm lim}-21))$. We take $a=4.56$ and $b=1$. $F(m)$ simply denotes a conversion from magnitude to fluxes.
	We then draw the types and redshifts from the multidimensional histogram shown in the other panels of \figref{fig:tdhistogram}.
	As emphasized previously, the latter is a binned version of the continuous prior distributions implemented in BPZ.
	For each object, we use the spectral template and the filters from \figref{fig:bandsandtemplates} to compute the other four magnitudes $ugrz$, and also add noise. 
	 These steps provide us with a set of noisy photometric fluxes following the distributions of interest. 
	 We now demonstrate that using the method presented above we recover the input distributions as well as the individual redshifts and types of galaxies in this simulated photometric catalogue.

	 We run the Gibbs sampler\footnote{The sampler is initialized at a raw histogram of the maximum likelihood redshifts of the individual objects. In all our tests the sampler converged very quickly and the initialization did not impact the final results.} and obtain samples from the posterior distribution $p(\{ t_\gal, z_\gal, m_\gal\}, \{f_{\ti\zi\mi}\} \ | \ \{ \hat{F}_{b,\gal}\} )$. Drawing $10^4$ samples takes about ten minutes using a laptop equipped with a 2.8 GHz Intel Core i7 processor. 
	 
	 \figrefbegin{fig:recovereddistributions} shows the distributions of type and redshift (with the reference magnitude marginalized over) obtained with samples $\{f_{\ti\zi\mi}\}$ from the full posterior distribution. 
	 For comparison, the stacked likelihood functions of all objects (summed over types and magnitudes) are also shown. 
	 The redshift distributions of the three types are successfully recovered with meaningful uncertainties. 
	 As observed in this figure and confirmed by the left panels of \figref{fig:objlikeandposts}, the likelihoods distributions are broad and multimodal, demonstrating that stacking them does not yield a meaningful estimate of the distributions of interest. 
	 This is a well-known drawback of template-fitting photometric redshift estimation methods: strong distribution priors $p(z,t,m)$ are usually required to deliver reliable redshift estimates \citep[see \eg][]{Benjamin:2012qp, Hildebrandt:2011hb, Bonnett:2015pww}.  
	 This is naturally alleviated in our Bayesian hierarchical model, and the mechanism by which the secondary peaks are suppressed is illustrated in a simplified setting in Appendix~\ref{section:shrinkage}.
	 
	 \figrefbegin{fig:dirichletparamsposterior} offers a closer look at the posterior distributions on the overall
	 fractional counts in the redshift bins of the model, \ie on the parameters $g_\zi = \sum_{\ti\mi} f_{\ti\zi\mi}$ which describe the overall redshift distribution of the sample. 
	 As expected, adjacent bins are significantly (anti)correlated, and correlation strength decreases with the separation between bins. 
	 Anticorrelation is a feature of the Dirichlet model but is significantly enhanced in our model due to the large support of the individual likelihoods. 
	 For comparison, the red contours show a noise-free inference, \ie samples from a simple Dirichlet model where the number counts are calculated from the true redshifts, types and magnitudes of our simulated catalogue. 
	 The contours are much smaller and exhibit smaller correlations, supporting the previous argument. 

	\figrefbegin{fig:objlikeandposts} shows the result of the our parameter inference on the redshift and type estimates of three randomly chosen individual objects (one per type). 
	The left panels highlight the typical problem of redshift estimation using photometric data: the likelihood functions are multimodal in both redshift and type. 
	The right panels show the maximum a posteriori distributions on type and redshift resulting from constraining the parameters of our hierarchical model. 
	Some of the degeneracies are removed; for example, secondary lobes at high redshift, which are clearly disfavoured by the inferred distributions shown in \figref{fig:recovereddistributions}, have been removed. 
	This demonstrates that the uncertainties in the individual redshifts and types are significantly reduced by including distribution information.
	Importantly, this new method inferred these redshift distributions directly from the data and propagated their uncertainties into the individual redshift and type estimates.
	This feature of the hierarchical models is sometimes referred to as `Bayesian shrinkage' as explained in more detail in Appendix~\ref{section:shrinkage}.

\section{Conclusions}\label{sec:conclusion}
	
	Current methods for estimating redshift distributions rely on stacking the posterior probability distributions on the individual redshifts of galaxies.
	These are typically obtained via template-fitting or machine learning techniques, and are sensitive to the available priors and training sets \citep[\eg][]{Hildebrandt:2011hb, Dahlen:2013fea, Bonnett:2015pww}.	
	The method we presented in this paper is the first to simultaneously infer the redshift distributions and individual redshifts of galaxies from noisy photometric flux measurements. 
	This new approach can be seen as a generalization of template-fitting methods and has several advantages over existing frameworks. 
	Simultaneously inferring individual redshifts and redshift distributions alleviates the sensitivity to external priors.
	In addition, the histogram representation of the distributions is very flexible and can represent a range of distributions, capturing features comparable to the bin size. 
	Its relation to the Dirichlet distribution allowed us to design a Gibbs sampling scheme to efficiently infer all the parameters of the model.
	Any library of spectral templates could be used in this framework to model the flux-redshift relation.
	In fact, multiple libraries could be used jointly since any degeneracies between the parameters will be inferred correctly, even with highly correlated templates. 
	We have also shown that the inclusion of selection effects such as redshift tomography cuts only affects the likelihood function, not the inference technique. 
	
	Our technique provides samples of the posterior distributions on redshift distributions. 
	These samples can be jointly used with galaxy clustering or cosmic shear  likelihood functions. 
	Therefore, our method is the first to consistently exploit spectral templates and provide a way to self-consistently propagate redshift uncertainties in cosmological analyses, which is essential for obtaining meaningful results from ongoing and upcoming photometric surveys.
	
	Our hierarchical model can be extended in various ways to include other physical effects of interest.
	It could exploit more complicated likelihood functions, for example including combinations of templates and marginalization of extra nuisance parameters such as magnitude zero points.
	It could also be interfaced with existing probabilistic approaches for analysing galaxy surveys, such as Bayesian methods for reconstructing the matter density field \citep[\eg][]{Jasche:2009hz, Jasche:2012kq} and inferring galaxy clustering or cosmic shear power spectra  \citep[\eg][]{Jasche:2013lwa, Alsing:2015zca}, which do not typically infer redshifts from noisy photometric fluxes.
	Together, these probabilistic techniques have the potential to unlock the true potential of galaxy surveys by extracting cosmological information while fully accounting for all uncertainties.

\section*{Acknowledgements}

BL thanks Alex I. Maltz and David Hogg for useful discussions. BL and HVP were supported by the European Research Council under the European Community's Seventh Framework Programme (FP7/2007- 2013) / ERC grant agreement no 306478-CosmicDawn. BL was also supported by the Simons Foundation and DJM by the STFC.


\bibliographystyle{mymnras_eprint}
\bibliography{bib}

\appendix
\section{Gaussian approximation}\label{app:gaussiancomp}

\begin{figure} 
\setlength{\unitlength}{1cm}
\centering\hspace*{-3mm}\includegraphics[width=8cm]{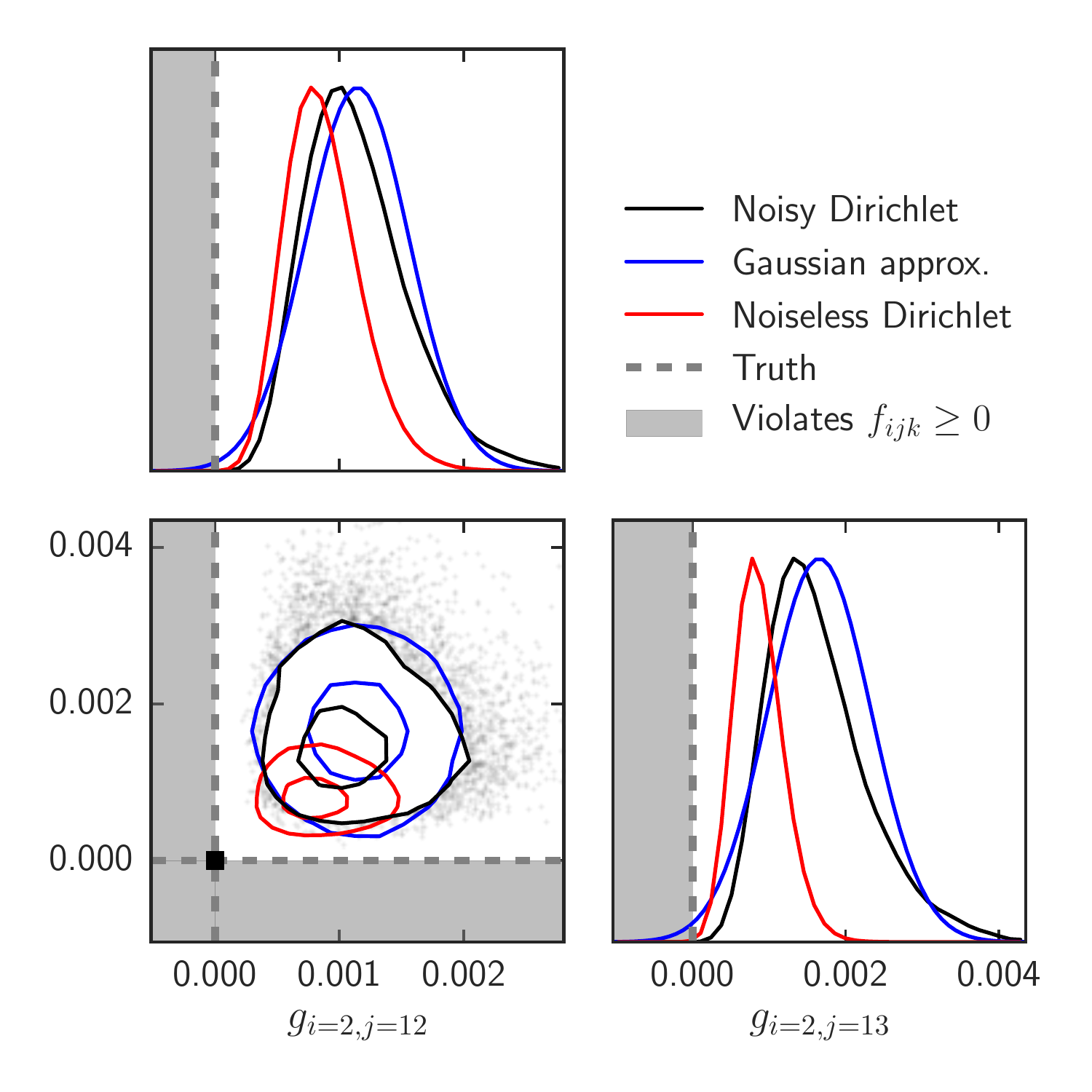}\vspace*{-3mm}
\caption{Posterior distributions on the parameters $g_{\ti\zi} = \sum_{\mi} f_{\ti\zi\mi}$ describing the second type-redshift distributions shown in \figref{fig:recovereddistributions}.}
\label{fig:posteriortyperedshiftdistributions}
\end{figure}

	One of the most common estimators for binned distributions and histograms assumes that the bins are uncorrelated and that the posterior distributions on the fractional count parameters are Gaussian. 
	For the parameters considered in this paper, this can be written as
	\equ{
		p(\{f_{\ti\zi\mi}\}|\{n_{\ti\zi\mi}\}) = \prod_{\ti\zi\mi} p(f_{\ti\zi\mi}|n_{\ti\zi\mi}) = \prod_{\ti\zi\mi} \mathcal{N}\left(f_{\ti\zi\mi}; \frac{n_{\ti\zi\mi}}{\nobj}, \frac{n_{\ti\zi\mi}}{\nobj}\right).\label{eq:gaussianapprox}
	}
	This form is an approximation of the Dirichlet distribution, which is the correct posterior distribution for histograms. 
	This approximation is accurate in the limit of large, uncorrelated counts $n_{\ti\zi\mi}$.
	However, it fails to capture inter-bin correlations. 
	In addition, for low values of $n_{\ti\zi\mi}$ the posterior distribution yields negative values for $f_{\ti\zi\mi}$, which violates the model assumptions since the fractional counts $f_{\ti\zi\mi}$ must be positive and sum to $1$.
	We highlight these problems using the illustration of \secref{sec:demo}: 
	\figrefbegin{fig:posteriortyperedshiftdistributions} shows the posterior distributions on the last two parameters of the second type-redshift distributions of \figref{fig:recovereddistributions}.
	The correct Dirichlet posterior distributions in both the noiseless and noisy cases are strictly positive.
	They do not encompass the true values (which are very close to zero) which is a feature of the Dirichlet distribution in this regime.
	The blue lines of \figref{fig:posteriortyperedshiftdistributions} show the result of approximating the Dirichlet posterior distribution with a multivariate Gaussian distribution with the same mean and diagonal covariance (but no inter-bin correlations).
	This is similar to what would be obtained by using \equref{eq:gaussianapprox} in place of the Dirichlet distribution in our hierarchical model applied to noisy photometric data.
	This approximation significantly goes to negative values of $f_{\ti\zi\mi}$ and fails to capture the inter-bin correlations.
	This highlights the need to employ the correct Dirichlet distribution when inferring the parameters of a binned distribution or a histogram.

\section{Suppression of secondary redshift peaks}
\label{section:shrinkage}

\begin{figure*} 
\includegraphics[width=6cm]{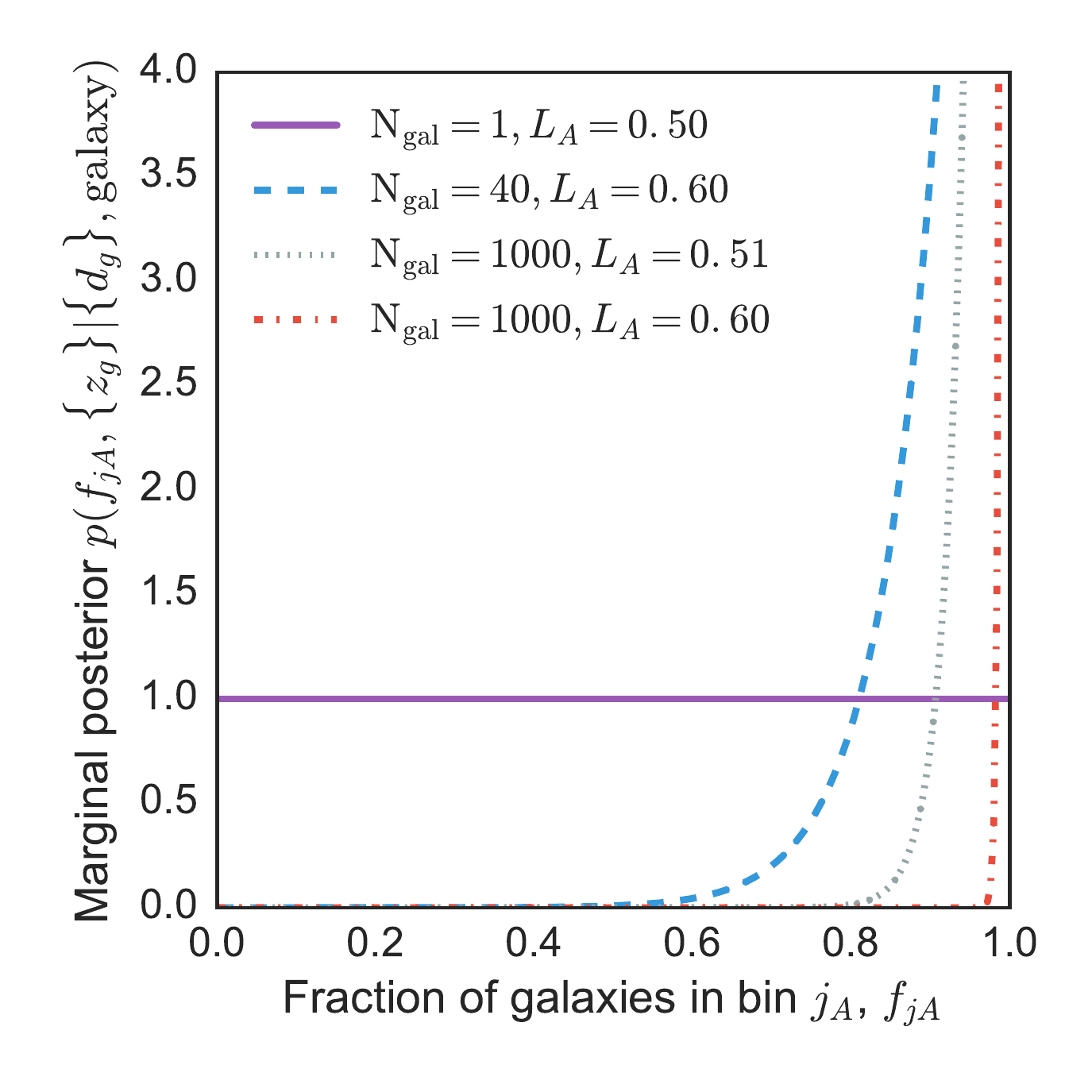}
\includegraphics[width=6cm]{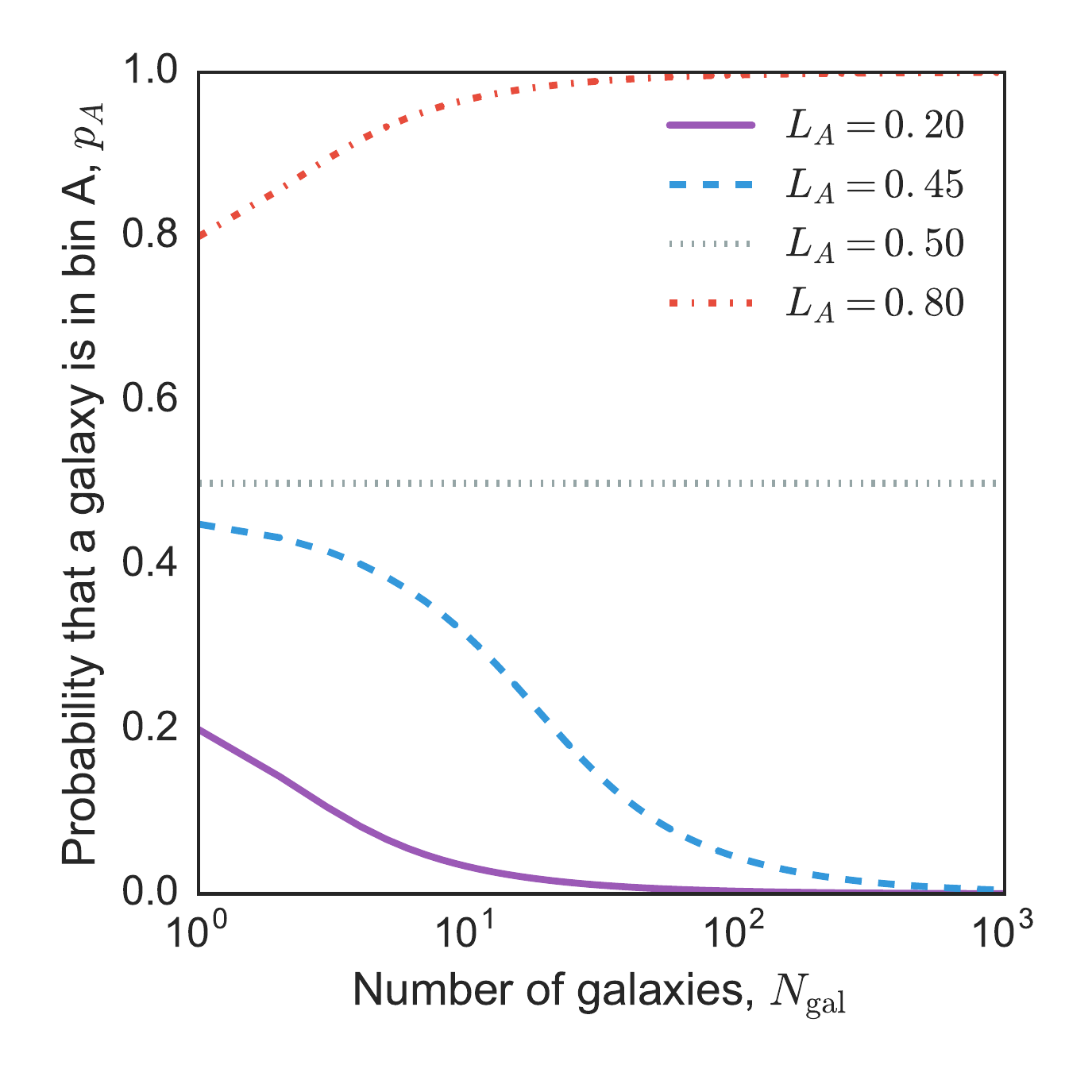}
\vspace*{-3mm}
\caption{Left: Posterior distribution on $f_{j_A}$ given in \equref{eq:posttoy_f} for various values of $\nobj$ and $L_A$ (with $f_{jB} = 1- f_{j_A}$ and $L_B = 1 - L_A$). 
Right: Value of $p_A$, the probability that a given galaxy is at $z_A$.
These highlight how small deviations from $L_A = L_B$ in the likelihood of individual galaxies are amplified in the posterior distributions on $p_A$ and $p_B$ and on the population parameters $f_{j_A}$ and $f_{j_B}$.
This is an illustration of the `(Bayesian) shrinkage' that occurs in the context of hierarchical models.}
\label{fig:toy}
\end{figure*}

One important aspect of the method presented here for inferring
redshift distributions is 
that the secondary peaks in the likelihood of some galaxies are 
strongly suppressed in the final inferred redshift distribution.
This is seen most clearly in the left-hand panel of 
\figref{fig:recovereddistributions}:
the peak in the stacked likelihoods at
$z \simeq 2.4$ is completely absent in the 
redshift distribution
as obtained from the Bayesian hierarchical model.
The complete absence of any peak associated with these 
high likelihoods is perhaps counterintuitive --- 
a model with two differently-weighted redshift peaks would 
surely be reasonable?
---
but the process by which these peaks are suppressed can be seen
explicitly by considering a greatly simplified version of the full
problem.

The aim, as in \secref{sec:popmodel}, is to infer the 
redshift distribution, $p(z | {\rm galaxy})$
of a population of galaxies,
but here only from a restricted sample of 
galaxies with strongly bimodal likelihoods.
Following \equref{eq:model_pwc}, 
the redshift distribution is modelled as being
piece-wise constant, 
so that
\begin{equation}
p(z | {\rm galaxy}) 
  = \sum_{j = 1}^{M}
  \frac
  {\Theta(z - z_{j,{\rm min}}) \, \Theta(z_{j,{\rm max}} - z)}
  {z_{j,{\rm max}} - z_{j,{\rm min}}} 
  f_j,
\end{equation}
where $z_{j+1,{\rm min}} = z_{j,{\rm max}}$
for $1 \leq j \leq M - 1$
(\ie the bins are contiguous)
and $f_j$ is the fraction of galaxies in the $j$'th bin.

The data consist of a sample of $\nobj$ very similar galaxies,
indexed by $g \in \{1, 2, \ldots, \nobj \}$,
with photometric data 
$d_g$ that in all cases leads to a bimodal likelihood of the form
\begin{equation}
\label{eq:toylik}
p(d_g | z_g) 
  = L_A \, \delta_\textrm{D}(z - z_A) + L_B \, \delta_\textrm{D}(z - z_B),
\end{equation}
where 
$L_A$, $L_B$, $z_A$ and $z_B$ all determined by $d_g$.
In other words, the photometric data for all of these $\nobj$ 
galaxies are consistent with one of two redshifts,
much as is the case for the galaxy whose likelihood is 
plotted in the top-left panel of \figref{fig:objlikeandposts}.

These galaxies can only contribute to the two bins
$j_A$ 
(for which $z_{j_A,{\rm min}} \leq z_A \leq z_{j_A,{\rm max}}$)
and $j_B$
(for which $z_{j_B,{\rm min}} \leq z_B \leq z_{j_B,{\rm max}}$); 
it is assumed that the two redshifts are sufficiently 
separated that $j_A \neq j_B$.
Hence in analysing this sample of $\nobj$ galaxies 
$f_j$ will be zero for all the other bins
and so it is only 
$f_{j_A}$ and $f_{j_B}$ which need to be inferred here.
Moreover, 
$f_{j_B} = 1 - f_{j_A}$, so that 
$f_{j_A}$ is the only independent population-level parameter.

Taking the prior on 
$f_{j_A}$ as uniform 
between 0 and 1
(a special case of the Dirichlet distribution used as the 
prior in Eq.~\ref{eq:prior}), the joint
prior on $f_{j_A}$ and the $\nobj$ redshifts $\{z_g\}$ of the galaxies is
\eqn{
p(f_{j_A}, \{z_g\} | {\rm galaxy})
  & = & \Theta(f_{j_A}) \, \Theta(1 - f_{j_A})
  \\
& \times &
  \prod_{g = 1}^\nobj
  \left[
  \frac
  {\Theta(z_g - z_{j_A,{\rm min}}) \, \Theta(z_{j_A,{\rm max}} - z_g)}
  {z_{j_A,{\rm max}} - z_{j_A,{\rm min}}}
  f_{j_A}
\right.
\nonumber \\
\label{eq:toyprior}
  & & + 
  \left.
  \frac
  {\Theta(z_g - z_{j_B,{\rm min}}) \, \Theta(z_{j_B,{\rm max}} - z_g)}
  {z_{j_B,{\rm max}} - z_{j_B,{\rm min}}}
  (1 - f_{j_A})
  \right].
\nonumber
}
Combining this prior with the likelihood given in \equref{eq:toylik}
leads to the 
full posterior 
\begin{equation}
p(f_{j_A}, \{z_g\} | \{d_g\}, {\rm galaxy})
 =
  \Theta(f_{j_A}) \, \Theta(1 - f_{j_A})
\end{equation}
\[
\,\,
  \times \,
(\nobj + 1)
\frac{
  \prod_{g = 1}^\nobj
  \left[
  f_{j_A} \, L_A \, \delta_\textrm{D}(z_g - z_A)
  + 
  (1 - f_{j_A}) \, L_B \, \delta_\textrm{D}(z_g - z_B)
  \right]
}
{
\sum_{g = 0}^\nobj
 L_A^g \, L_B^{\nobj - g}
}.
\]

Integrating the posterior over the $\nobj$ redshifts
yields the marginal posterior distribution in $f_{j_A}$ as
\begin{equation}
\label{eq:posttoy_f}
p(f_{j_A} | \{d_g\}, {\rm galaxy})
\end{equation}
\[
\,\,
  = \Theta(f_{j_A}) \, \Theta(1 - f_{j_A})
  \frac{
  \sum_{g = 0}^\nobj
  \frac{ (\nobj + 1)!}{g! \, (\nobj - g)!}
  (f_{j_A} \, L_A)^g \,
  [(1 - f_{j_A}) \, L_B]^{\nobj - g}
  }
  {\sum_{g = 0}^\nobj L_A^g \, L_B^{\nobj-g}
  }.
\]
Some examples of this posterior 
for different $\nobj$, $L_A$ and $L_B$ are shown in 
the left-hand panel of \figref{fig:toy}.
If $L_A = L_B$ 
this posterior is uniform between $f_{j_A} = 0$ and 
$f_{j_A} = 1$, which is the same as the prior -- 
the data provide no reason to prefer either option, 
no matter how many galaxies are in the sample.
Otherwise, taking $L_A > L_B$ (without loss of generality, as the
labels are arbitrary), 
the posterior mode is $f_{j_A} = 1$, even with just one galaxy in the sample.
The posterior becomes increasingly sharply peaked 
with increasing $\nobj$ even if $L_A$ is only marginally greater than
$L_B$.
Provided only that $L_A > L_B$,
$\lim_{\nobj \rightarrow \infty} p(f_{j_A} | \{d_g\}, {\rm galaxy}) 
  = \delta_\textrm{D}(f_{j_A} - 1)$.

One result of the potentially sharp posterior in $f_{j_A}$ for
large $\nobj$ is that 
the posterior 
distributions of the redshift of the individual galaxies
(which are all the same) can be 
very different from their associated likelihoods.
Integrating over $f_{j_A}$ and 
$\nobj - 1$ galaxy redshifts
(and hence assuming that $\nobj \geq 1$)
yields the marginal posterior distribution on the redshift 
of any one of the $\nobj$ galaxies as
\begin{equation}
p(z_g | \{d_g\}, {\rm galaxy})
  = 
  p_A \, \delta_\textrm{D}(z_g - z_A)
  +
  p_B \, \delta_\textrm{D}(z_g - z_B),
\label{eq:posttoy_z}
\end{equation}
where 
\begin{equation}
p_A = 
\frac{
\sum_{g = 0}^{\nobj - 1} (g + 1) \, L_A^{g + 1} \, L_B^{\nobj - 1 - g}
}
{
\nobj \sum_{g = 0}^\nobj L_A^g \, L_B^{\nobj - g}
}
\end{equation}
and
\begin{equation}
p_B = 
\frac{
\sum_{g = 0}^{\nobj - 1} (g + 1) \, L_A^{\nobj -1 - g} \, L_B^{g + 1}
}
{
\nobj \sum_{g = 0}^\nobj L_A^g \, L_B^{\nobj - g}
}
\end{equation}
are,
respectively, the probabilities that the galaxy is at $z_A$ and $z_B$.
In the case that $\nobj = 1$ this result reduces to
$p_A = L_A / (L_A + L_B)$
and 
$p_B = L_B / (L_A + L_B)$,
and if $L_A = L_B$ then
$p_A = p_B = 1/2$.
Otherwise (and once again adopting $L_A > L_B$),
$p_A$ increases with $L_A$ and $\nobj$,
as illustrated in the right-hand panel of \figref{fig:toy}.
Of particular interest is the case that $L_A$ is only a little 
higher than $L_B$ but $\nobj$ is high: 
the galaxy is almost certainly at $z_B$, despite the observed data
on this particular object being ambiguous.
This is an example of `borrowing strength' or `(Bayesian) shrinkage',
in which most of the information about an individual object comes 
from the population of which it is a member.




\end{document}